# Comparison of Gini index and Tamura coefficient for holographic autofocusing based on the edge sparsity of the complex optical wavefront

**Authors:**
Miu Tamamitsu[1,2,3]†, Yibo Zhang[1,2,3]†, Hongda Wang[1,2,3], Yichen Wu[1,2,3], and Aydogan Ozcan[1,2,3,4,*]

**Affiliations:**
[1]Electrical and Computer Engineering Department, University of California, Los Angeles, CA, 90095, USA.
[2]Bioengineering Department, University of California, Los Angeles, CA, 90095, USA.
[3]California NanoSystems Institute (CNSI), University of California, Los Angeles, CA, 90095, USA.
[4]Department of Surgery, David Geffen School of Medicine, University of California, Los Angeles, CA, 90095, USA.

[*]Correspondence: Prof. Aydogan Ozcan
E-mail: ozcan@ucla.edu
Web: http://innovate.ee.ucla.edu/welcome.html
Address: 420 Westwood Plaza, Engr. IV 68-119, UCLA, Los Angeles, CA 90095, USA.
Tel: +1(310)825-0915
Fax: +1(310)206-4685
†Equal contributing authors

**Authors' email addresses:**
Miu Tamamitsu: tamamitsu@ucla.edu
Yibo Zhang: zybmax@ucla.edu
Hongda Wang: hdwang@ucla.edu
Yichen Wu: wuyichen@ucla.edu
Aydogan Ozcan: ozcan@ucla.edu

**Abstract:**
The Sparsity of the Gradient (SoG) is a robust autofocusing criterion for holography, where the gradient modulus of the complex refocused hologram is calculated, on which a sparsity metric is applied. Here, we compare two different choices of sparsity metrics used in SoG, specifically, the Gini index (GI) and the Tamura coefficient (TC), for holographic autofocusing on dense/connected or sparse samples. We provide a theoretical analysis predicting that for uniformly distributed image data, TC and GI exhibit similar behavior, while for naturally sparse images containing few high-valued signal entries and many low-valued noisy background pixels, TC is more sensitive to distribution changes in the signal and more resistive to background noise. These predictions are also confirmed by experimental results using SoG-based holographic autofocusing on dense and connected samples (such as stained breast tissue sections) as well as highly sparse samples (such as isolated *Giardia lamblia* cysts). Through these experiments, we found that ToG and GoG offer almost identical autofocusing performance on dense and connected samples, whereas for naturally sparse samples, GoG should be calculated on a relatively small region of interest (ROI) closely surrounding the object, while ToG offers more flexibility in choosing a larger ROI containing more background pixels.

**Introduction**

In digital holographic imaging, accurate estimation of a specimen's focus distance ("*z* distance") is essential to create accurate reconstructions [1-4]. Automatic determination of the focus distance (i.e., autofocusing) is a widely researched topic with numerous solutions [5-12]. A commonly used strategy is to design a function, called an autofocusing criterion, that quantifies how "focused" an image is. Then, this function is evaluated on digitally refocused holographic images at various distances, and the distance that corresponds to the maximum (or minimum) of this function is chosen as the focus distance. However, various previously proposed autofocusing methods suffer from either lack of generality for specimens of different optical transmission functions (i.e., amplitude-contrast, phase-contrast, or mixed) [5], changing polarities (peak or valley at the correct *z*) for different object types [5], or lack of robustness to strong twin image artifacts when autofocusing on dense and connected specimens. These limitations make unsupervised autofocusing rather challenging. Recently, we have found that the edge sparsity of a refocused complex-valued image can be used as a robust autofocusing criterion on a wide variety of samples including dense and connected specimens as well as objects with different levels of amplitude and/or phase-contrast compositions. The edge sparsity constraint can be quantified using a measure that we introduce, i.e., the Sparsity of the Gradient (SoG), given by:

$$\mathrm{SoG}(U) = \mathrm{S}(|\nabla U|) \tag{1}$$

where $U$ is the complex refocused image at a certain distance, $\nabla$ is the gradient operator, $|\cdot|$ is the modulus, and $\mathrm{S}(\cdot)$ represents a sparsity metric. $|\nabla U|$ can be approximated for a discrete complex-valued image as:

$$|\nabla U|_{i,j} = \left( |U_{i,j} - U_{i,j-1}|^2 + |U_{i,j} - U_{i-1,j}|^2 \right)^{\frac{1}{2}} \tag{2}$$

There are various choices for the sparsity metric, S. However, among the ones summarized in Ref. [13], only the Gini index (GI) and *pq*-mean possess all the intuitive attributes that a sparsity metric should have. Therefore, here we specifically compare the performances of GI and an equivalent form of *pq*-mean when *p* = 1 and *q* = 2 (i.e., the Tamura coefficient, TC [9]) when used as a sparsity metric in SoG.

**Theoretical comparison of Tamura coefficient and Gini index as a sparsity metric**

In this section, we give a theoretical analysis of TC and GI and report their performance as a sparsity metric. It has been shown that both GI and *pq*-mean satisfy the six desirable attributes of a sparsity metric and hence can be used to represent the sparsity of data [13]. In this section, we first show that TC is a monotonic transformation of *pq*-mean with *p* = 1 and *q* = 2 and therefore, also satisfies the same six desirable attributes [13] to serve as a sparsity metric. Then, we investigate the mathematical formulas of TC and GI to reveal their fundamental differences from the perspective of holographic autofocusing. *Specifically, when the image pixel data are uniformly distributed, the two metrics show similar SoG-based autofocusing behavior; but when the image is naturally sparse, GI tends to become less sensitive to distribution changes in high-valued entries of the image data and more sensitive to the low-valued entries, unlike TC.*

The definition of *pq*-mean is given by:

$$\tilde{\mathrm{M}}_{p,q}(\mathbf{c}) = -\mathrm{M}_{p,q}(\mathbf{c}) = -\frac{\left(\frac{1}{N}\sum_{j=1}^{N} c_j^p\right)^{\frac{1}{p}}}{\left(\frac{1}{N}\sum_{j=1}^{N} c_j^q\right)^{\frac{1}{q}}} \tag{3}$$

which is simply the negative ratio of the generalized *p*-mean and *q*-mean, where **c** is the data vector with $c_j \geq 0$ (*j* = 1, 2, ..., *N*) and it is required that *p* ≤ 1 and *q* > 1 for *pq*-mean to serve as a sparsity metric [13]. **c** cannot be a zero vector. The definition of TC is given by:

$$\text{TC}(\mathbf{c}) = \sqrt{\frac{\sigma(\mathbf{c})}{\langle \mathbf{c} \rangle}} \qquad (4)$$

where σ(·) is the standard deviation, and $\langle \cdot \rangle$ refers to the mean. Based on these definitions, one can show that:

$$\text{TC}(\mathbf{c}) = \sqrt{\frac{\left(\frac{1}{N}\sum_{j=1}^{N}\left(c_j - \frac{1}{N}\sum_{j=1}^{N}c_j\right)^2\right)^{\frac{1}{2}}}{\frac{1}{N}\sum_{j=1}^{N}c_j}} = \left[\frac{\frac{1}{N}\sum_{j=1}^{N}c_j^2}{\left(\frac{1}{N}\sum_{j=1}^{N}c_j\right)^2} - 1\right]^{\frac{1}{4}} = \left[\tilde{M}_{1,2}^{-2}(\mathbf{c}) - 1\right]^{\frac{1}{4}} \qquad (5)$$

Equation (5) indicates that TC(**c**) is a monotonical transformation of $\tilde{M}_{1,2}(\mathbf{c})$ and since $\tilde{M}_{1,2}(\mathbf{c})$ has already been shown to satisfy the six desirable attributes of a sparsity metric [13], it is clear that TC(**c**) also does.

Next, we compare these two metrics to shed more light on their differences. The definition of GI [7] is

$$\text{GI}(\mathbf{c}) = 1 - 2\sum_{k=1}^{N}\frac{a_{[k]}}{\text{sum}(\mathbf{c})}\left(\frac{N-k+0.5}{N}\right) = 1 - \frac{2}{N}\sum_{k=1}^{N}\tilde{a}_{[k]}\left(\frac{N-k+0.5}{N}\right) = \frac{2}{N}\sum_{k=1}^{N}\frac{k}{N}\tilde{a}_{[k]} - 1 - \frac{1}{N} \qquad (6)$$

where $a_{[k]}$ ($k = 1, ..., N$) are the sorted entries of **c** in ascending order, and $\tilde{a}_{[k]} = \frac{a_{[k]}}{\langle \mathbf{c} \rangle}$ are the sorted entries of **c** normalized by the mean of **c**. TC can be also written as:

$$\text{TC}(\mathbf{c}) = \left(\frac{1}{N}\sum_{k=1}^{N}\tilde{a}_{[k]}^2 - 1\right)^{\frac{1}{4}} \qquad (7)$$

in which the entries are sorted and normalized similarly, without changing the calculation result. If we compare the essential parts of Eq. (6) and (7), i.e., $\sum_{k=1}^{N}\frac{k}{N}\cdot\tilde{a}_{[k]}$ and $\sum_{k=1}^{N}\tilde{a}_{[k]}\cdot\tilde{a}_{[k]}$, respectively, one can see that the mechanism through which each metric measures the sparsity of data is based on a weighted sum of the entries, where larger weights are given to larger entries and smaller weights are given to smaller entries. For GI, the weight given to each entry is proportional to the ranking of the entry in ascending order, while for TC, the weight is proportional to the entry's value itself.

This difference in the weight given to each data entry in TC and GI makes their behavior strongly dependent on the statistical distribution of the image data. One condition for the two to behave similarly is when the data entries are almost uniformly distributed between 0 and a positive number so that $\tilde{a}_{[k]}$ is roughly proportional to $\frac{k}{N}$, making the pixel weights in TC and GI similar to each other. A condition for the two to behave rather different is when the image is highly sparse. For example, when only a few high-valued entries of the image data form the signal of interest and the remaining low-valued entries form the background noise, then the different weights in the definitions of TC and GI lead to two effects. First, the weight that GI gives to high-valued entries (the "signal") becomes $\frac{k}{N} \sim 1$, resulting in less discrimination within the "signal" values. *This makes GI insensitive to the changes in the distribution of high-valued signal entries in a sparse sample, unlike TC, which reflects the value of the entry itself* *to its weight.* Second, *assuming the background noise level is much lower than the sparse signal strength, GI gives higher relative weight to the noisy background entries than TC does, thus is more susceptible to noise, especially in sparse samples. Because of this, it is expected that TC can tolerate more background noise than GI does, when the signal of interest is sparse.* These predictions will be tested experimentally in the next section.

**Experimental comparison of Tamura coefficient and Gini index for holographic autofocusing based on edge sparsity of the complex optical wavefront**

In this section, we give an experimental comparison of the performance of TC and GI as a sparsity metric used in SoG for holographic autofocusing on *naturally sparse* as well as *dense and connected* samples. We term the autofocusing algorithm based on TC of the gradient modulus as **ToG**, and that based on GI of the gradient modulus as **GoG**.

To investigate the performance of ToG and GoG for *naturally sparse samples*, we used an in-line hologram of an isolated *Giardia lamblia* cyst (a waterborne parasite) and performed holographic autofocusing on it (see Fig. 1). Both non-phase-retrieved and phase-retrieved scenarios are considered in Fig. 1, left and right panels, respectively. The phase retrieval (right panel) was performed through an iterative method using object support [14,15]. The in-line holograms as well as the in-focus reconstructed images are shown in Fig. 1 (a, c, e and g). Their corresponding spatial gradient moduli are also shown in Fig. 1 (b, d, f and h). As is obvious in Fig. 1(d and h), only a small portion of each image of the spatial gradient modulus is occupied by the signal from the object, and hence these spatial gradient moduli are naturally sparse.

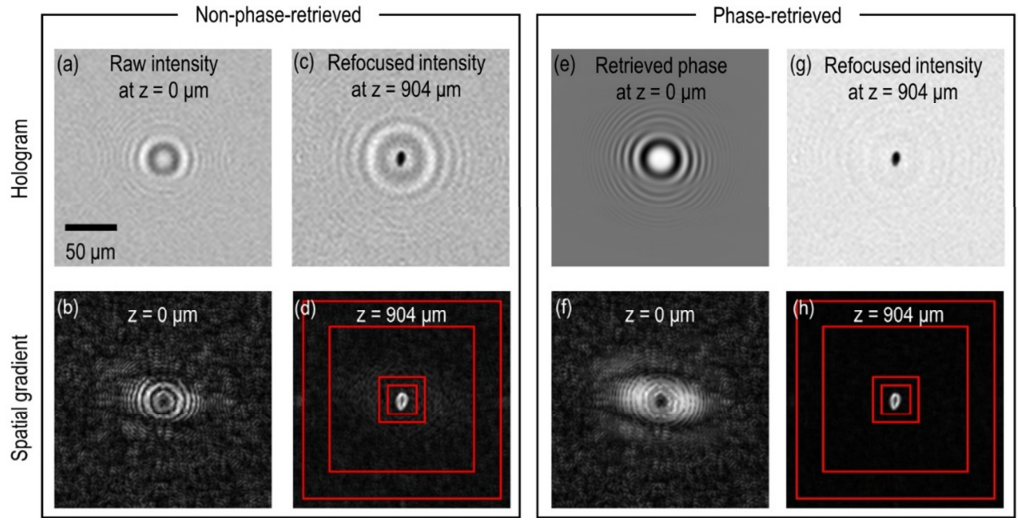

**Fig. 1**. Hologram of a *Giardia lamblia* cyst and the spatial gradient modulus of the corresponding complex field at different focus planes. (Left) Non-phase-retrieved results. (Right) Phase-retrieved results. (a) Raw hologram of the *Giardia lamblia* cyst at $z$ = 0 μm, (b) its spatial gradient modulus, (c) in-focus reconstruction of the hologram at $z$ = 904 μm and (d) the spatial gradient modulus of the reconstructed complex field at $z$ = 904 μm. (e) Retrieved phase of the raw hologram at $z$ = 0 μm, (f) the spatial gradient modulus of the phase-retrieved complex hologram, (g) in-focus reconstruction of the phase-retrieved hologram at $z$ = 904 μm and (d) the spatial gradient modulus of the reconstructed complex field at $z$ = 904 μm. Red squares shown in the images of the spatial gradient modulus (d and h) indicate the cropped ROIs within which TC and GI are calculated in the autofocusing process. The sizes of the red squares are 28 μm × 28 μm, 45 μm × 45 μm, 140 μm × 140 μm and 190 μm × 190 μm, respectively, with a pixel pitch of 1.4 μm.

We performed wave propagation of the hologram with different distances ranging between 600 - 1200 μm with a step size of 1 μm and calculated the corresponding spatial gradient modulus at each distance. We cropped a certain ROI out of each of the obtained spatial gradient modulus (Fig. 1 (d and h)) and calculated the sparsity metrics on the cropped ROI as a function of the focus distance (see Fig. 2). The focus distance that corresponds to the maximum sparsity value was chosen as the autofocusing result for ToG and GoG, respectively. As evident in Fig. 1, the non-phase-retrieved holograms are corrupted by the twin-image artifact, representing a much more challenging scenario for autofocusing than its phase-retrieved counterpart.

Figure 2 shows the holographic autofocusing results for the *Giardia lamblia* cyst shown in Fig. 1. ToG and GoG with ROI sizes of (a and e) 28 μm × 28 μm, (b and f) 45 μm × 45 μm, (c and g) 140 μm × 140 μm and (d and h) 190 μm × 190 μm were calculated for different focus distances. The refocused intensity image corresponding to ToG or GoG's maximum point is also shown for each case. We first discuss the non-phase-retrieved results shown in Fig. 2(a-d). For the 28 μm × 28 μm ROI, ToG and GoG show a similar trend in their autofocusing results: the reconstructed intensity images obtained by using ToG and GoG are both in-focus. However, if we increase the ROI size to 45 μm × 45 μm or 140 μm × 140 μm, GoG starts to fail while ToG still succeeds to autofocus, as shown in Fig. 2(b and c). With an ROI size of 190 μm × 190 μm, ToG is also affected by the larger amount of noise in the metric curve and fails to find the correct focus. Now, if we advert to the phase-retrieved results shown in Fig. 2(e-h), the metric curves of ToG and GoG are smoother and sharper than their non-phase-retrieved counterparts due to the suppressed twin-image artifact, as expected. While GoG is still not successful in autofocusing with a ROI larger than 45 μm × 45 μm, ToG finds the correct focus even with a ROI of 190 μm × 190 μm centered at the cyst location. These results confirm our previous conclusion: TC's higher sensitivity to high-valued signal entries and better resistivity to low-valued noise terms make ToG's ROI selection more flexible than GoG for sparse samples. ***This suggests that the users should be careful not to choose too large ROIs around an object, especially when the sample is extremely sparse.***

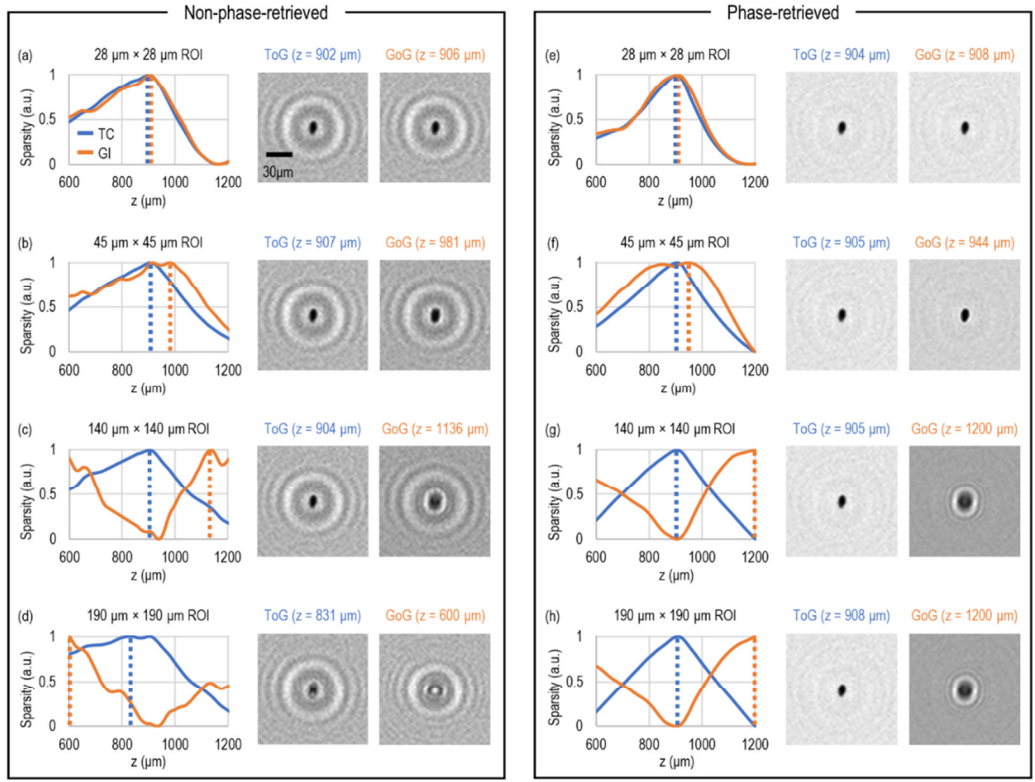

**Fig. 2**. Performance comparison between ToG and GoG for a sparse sample, based on (left) the non-phase-retrieved raw hologram and (right) the phase-retrieved complex hologram. The autofocusing was performed on different ROI sizes of (a and e) 28 μm × 28 μm, (b and f) 45 μm × 45 μm, (c and g) 140 μm × 140 μm and (d and h) 190 μm × 190 μm. In each figure panel, the normalized values of TC and GI are plotted, and the reconstructed intensity images of the object (a waterborne parasite cyst) based on ToG and GoG's resulting focus distances are shown to the right.

To better visualize and compare the behavior of TC and GI, in Fig. 3(c and f), we plotted the weights that are given to the pixels of the spatial gradient modulus image by TC and GI, calculated on a non-phase-retrieved *Giardia lamblia* cyst hologram and a non-phase-retrieved H&E-stained breast tissue sample hologram, refocused to the correct focus distances (Fig. 3(a, b, d and e)). Two ROI sizes were used in this comparison: 28 μm × 28 μm and 140 μm × 140 μm. To make the comparison easier, the x-axes of Fig. 3 (c and f) represent the pixel value of the spatial gradient moduli normalized by their respective maximum. Note that the normalization of the data does not change the GI and TC values, as they are scale-invariant [13]. The *y*-axes in these plots refer to the normalized weights of TC and GI. TC's weights (blue lines in Fig. 3 (c and f)) are shown by $y = x$ lines since TC's weight is proportional to the value of the pixel itself while GI's weight (yellow and orange curves in Fig. 3 (c and f)) is equal to the normalized ranking (i.e., the percentile rank) of a given pixel value of the spatial gradient modulus image, sorted in ascending order. Based on Fig. 3c, for a *Giardia lamblia* cyst (an isolated and sparse object), when the ROI size is 28 μm × 28 μm, a pixel with a value ranging between e.g., ~0.2 - 1 is given a normalized GI weight ranging between ~0.8 – 1, whereas when the ROI size is increased to 140 μm × 140 μm, the same range of pixel entries is given a GI weight between ~0.99 and 1. This major difference between GI and TC causes GI to be less sensitive to the distribution changes in high-valued entries for sparse data.

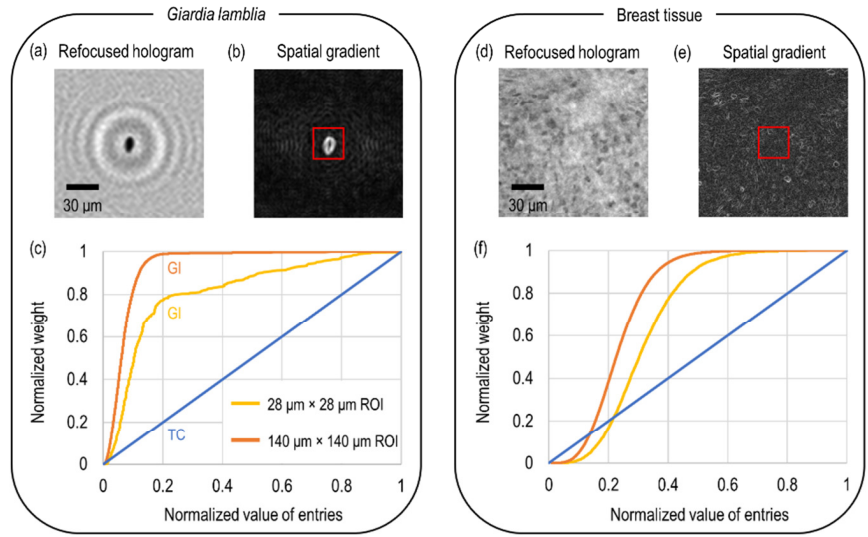

**Fig. 3**. Analysis of the effect of choosing different ROI sizes for sparsity calculations for (left) a *Giardia lamblia* cyst (representing a sparse sample) and (right) a breast tissue section (representing a dense, connected sample). (a and d) Refocused intensity images of the specimens and (b and e) the spatial gradient modulus of the corresponding refocused complex field. The shown images represent 140 μm × 140 μm ROI, whereas the red squares shown in the images of the spatial gradient modulus represent 28 μm × 28 μm ROI. (c and f) Normalized weights of TC and GI given to the pixels of the spatial gradient modulus, as a function of the normalized pixel value, calculated on 28 μm × 28 μm and 140 μm × 140 μm ROIs. TC's weight (blue line) is proportional to the pixel value itself, while GI's weight (yellow and orange curves) is equal to the normalized ranking (i.e., percentile rank) of the pixel value in the spatial gradient modulus image, in ascending order.

On the other hand, a non-sparse sample such as an H&E-stained breast tissue section (Fig. 3(d, f)) does not have the same issue. The data distribution does not change much when choosing different ROI sizes, due to the connected and dense nature of the sample. Moreover, because the data distribution of the breast specimen's spatial gradient modulus is less sparse and more uniform, it is expected that ToG and GoG should behave similarly. To validate this, we performed autofocusing using ToG and GoG on non-phase-retrieved holograms of breast tissue sections. The experimental results are summarized in Fig. 4, where two different parts of the same breast tissue are used, i.e., a region containing (a-d) more nuclei, named as "breast tissue (nuclei)" and (e-h) fewer nuclei, mainly stroma i.e., connected tissue, named as "breast tissue (stroma)". First, autofocusing on these two parts of the breast tissue was performed on relatively large ROIs of 300 μm × 300 μm (the entire regions shown in a and e, left panel), and the metric curves as a function of *z* are plotted in (b and f), left panel. As predicted, the metric curves are almost identical for GoG and ToG, both successfully identifying the correct focus distance at *z* = 377 μm. The non-phase-retrieved and phase-retrieved reconstructions at this plane are shown in (c, d, g and h). Next, we reduced the ROI sizes until we found the smallest ROI that can be successfully autofocused on, which, again, was identical for GoG and ToG: for the tissue section with more nuclei, it is 38 μm × 38 μm, and for the tissue section with more stroma, it is 75 μm × 75 μm, corresponding to the red squares in (a and e), respectively. We believe that the difference in the smallest "focusable" ROIs is due to different spatial features within these two areas: the nuclei might provide more distinct edges, boosting the autofocusing robustness. Furthermore, GoG and ToG's metric curves for the smaller ROIs also agree very well as shown in Fig. 4b, f. ***All these results suggest that for a planar, connected and dense object, the user can take advantage of a large ROI for better autofocusing accuracy using either ToG or GoG***.

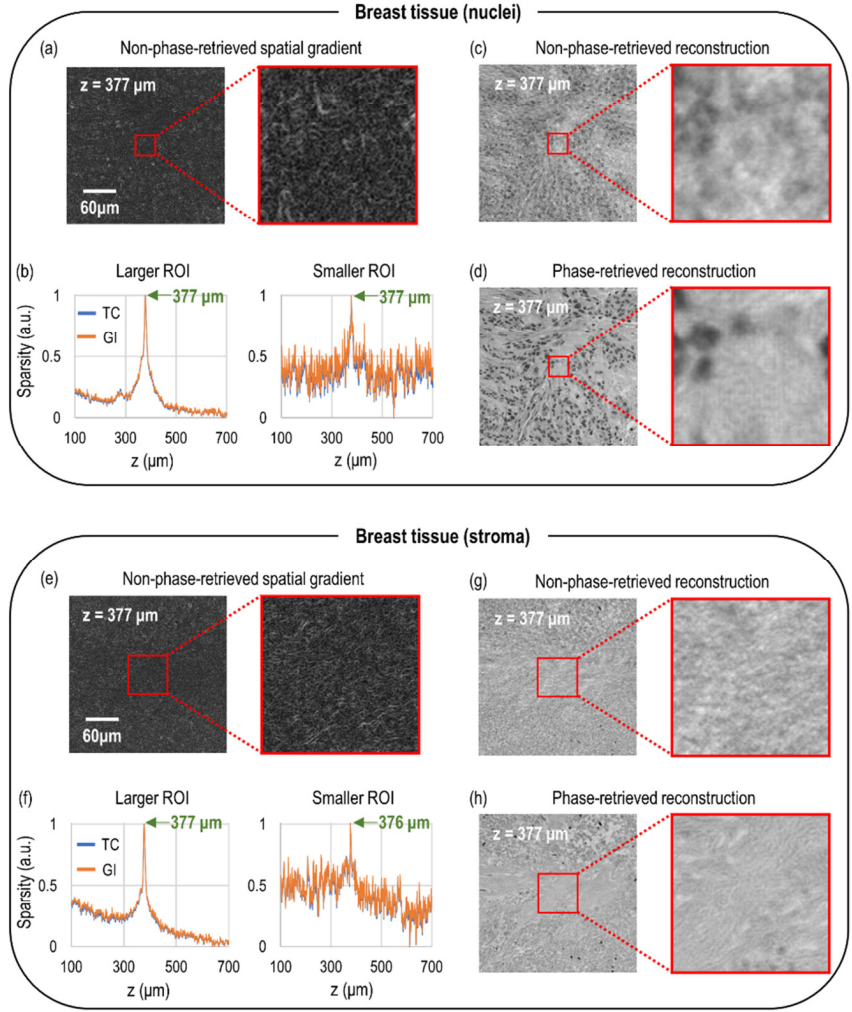

**Fig. 4**. Experimental results of ToG and GoG on dense and connected samples. Two different regions of the same breast tissue were used as the specimen, i.e., a region containing (a - d) more nuclei and (e - h) fewer nuclei. (a and e) Spatial gradient modulus of the in-focus complex field at z = 377 μm reconstructed from a non-phase-retrieved raw in-line hologram. The entire image of the spatial gradient modulus represents the larger ROI of 300 μm × 300 μm and the red squares shown in them represent smaller ROIs of (a) 38 μm × 38 μm and (e) 75 μm × 75 μm. (c and g) In-focus intensity image at z = 377 μm reconstructed from a non-phase-retrieved raw in-line hologram. (d and h) In-focus intensity image at z = 377 μm reconstructed from a phase-retrieved in-line hologram.

We also demonstrated that ToG and GoG are both applicable to a variety of samples. Shown in Fig. 5 are the results of ToG and GoG performed on different sparse samples: (a) a single bovine sperm, (b) a single yeast cell, and (c) two human red blood cells, with two ROI sizes for each of them: (a) 110 μm × 110 μm and 11 μm × 11 μm, (b) 170 μm × 170 μm and 17 μm × 17 μm, (c) 37 μm × 37 μm and 15 μm × 15 μm. Figure 5 once again demonstrates that for a sparse sample when the ROI is small, both ToG and GoG provide successful autofocusing on all the specimens. However, when the ROI around a sparse sample gets larger, GoG can result in out-of-focus images, whereas ToG still delivers in-focus images, as discussed in detail earlier.

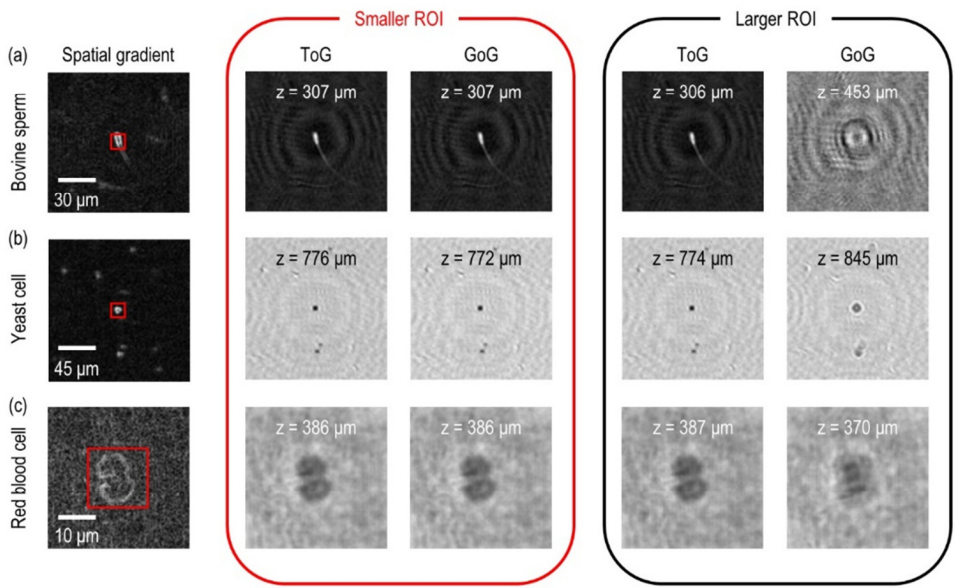

**Fig. 5**. Experimental results of ToG and GoG based holographic autofocusing on a variety of sparse samples such as (a) a bovine sperm cell, (b) a yeast cell and (c) two human red blood cells. The autofocusing was tested on two different ROI sizes for each sample: (a) 110 μm × 110 μm and 11 μm × 11 μm, (b) 170 μm × 170 μm and 17 μm × 17 μm, (c) 37 μm × 37 μm and 15 μm × 15 μm. The larger ROIs are represented by the images of the spatial gradient modulus, while the red square in each of them represents the corresponding smaller ROI.

Finally, in Table 1 we summarized autofocusing results on less sparse samples, including a USAF resolution test target, stained and unstained cell smears, and thin tissue sections, where the size of the in-line holograms were ~ 380 μm × 380 μm, and the ROIs were selected to be only slightly smaller than the hologram size (~ 24 μm cropped out from each side to avoid edge artifacts during coherent wave propagation). The absolute autofocusing errors (in μm) are reported in this table, which is defined as the absolute difference between the autofocused result and the ground truth (manual focusing). It is clear that for all the samples, very small errors (~ 1 μm) are achieved for both phase retrieved and non-phase retrieved holograms. These results also suggest that both GoG and ToG can also be applied to off-axis holographic imaging, where twin image does not pose a challenge.

| Phase retrieved? | No | | Yes | |
|---|---|---|---|---|
| Method | GoG | ToG | GoG | ToG |
| Polarity | Max | Max | Max | Max |
| USAF target | 0.96 | 1.02 | 0.52 | 0.88 |
| Unstained Pap Smear | 0.10 | 0.04 | 0.32 | 0.20 |
| Stain Pap Smear | 1.66 | 1.52 | 0.16 | 0.19 |
| Lung tissue | 0.12 | 0.54 | 0.04 | 0.30 |
| Blood smear | 1.08 | 1.10 | 0.76 | 0.85 |
| Breast tissue (nuclei) | 1.36 | 1.46 | 0.09 | 0.23 |
| Breast tissue (stroma) | 1.87 | 1.85 | 0.56 | 0.62 |
| Average | 1.02 | 1.08 | 0.35 | 0.47 |

**Table 1**. Comparison of the autofocusing results of GoG and ToG on less sparse samples, where the absolute errors, i.e., the absolute difference between the autofocusing result and the ground truth based on manual focusing are displayed (unit: μm).

**Acknowledgements**

The Ozcan Research Group at UCLA gratefully acknowledges the support of the Presidential Early Career Award for Scientists and Engineers (PECASE), the Army Research Office (ARO; W911NF-13-1-0419 and W911NF-13-1-0197), the ARO Life Sciences Division, the National Science Foundation (NSF) CBET Division Biophotonics Program, the NSF Emerging Frontiers in Research and Innovation (EFRI) Award, the NSF EAGER Award, NSF INSPIRE Award, NSF Partnerships for Innovation: Building Innovation Capacity (PFI:BIC) Program, Office of Naval Research (ONR), the National Institutes of Health (NIH), the Howard Hughes Medical Institute (HHMI), Vodafone Americas Foundation, the Mary Kay Foundation, Steven & Alexandra Cohen Foundation, and KAUST. This work is based upon research performed in a laboratory renovated by the National Science Foundation under Grant No. 0963183, which is an award funded under the American Recovery and Reinvestment Act of 2009 (ARRA).